\documentclass[aps,twocolumn]{revtex4}
\usepackage{amsmath,amssymb}

\usepackage{tabularx}

\usepackage{epsfig}
\usepackage{graphicx}

\newcommand{\be}{\begin{equation}}
\newcommand{\ee}{\end{equation}}
\newcommand{\bea}{\begin{eqnarray}}
\newcommand{\eea}{\end{eqnarray}}
\newcommand{\nn}{\nonumber}

\begin{document}

\title{Brane-bulk energy exchange : a model with the present universe as a global attractor}

\author{Georgios Kofinas$^1$\footnote{kofinas@ffn.ub.es},
Grigorios Panotopoulos$^2$\footnote{panotop@physics.uoc.gr} and
Theodore N. Tomaras$^{2,3}$\footnote{tomaras@physics.uoc.gr}}

\date{\today}

\address{~}

\address{$^{1}$Departament de F{\'\i}sica Fonamental,
Universitat de Barcelona, Diagonal 647, 08028 Barcelona, Spain}

\address{$^{2}$Department of Physics and Institute of Plasma Physics, University of Crete, 71003 Heraklion, Greece}

\address{$^{3}$Foundation of Research and Technology, Hellas, 71110 Heraklion, Greece}

\begin{abstract}

The role of brane-bulk energy exchange and of an induced gravity term on a single braneworld of negative
tension and vanishing effective cosmological constant is studied.
It is shown that for the physically interesting cases of dust and radiation
a unique global attractor which can realize our present universe
(accelerating and $0\!<\!\Omega_{m0}\!<\!1$) exists for a wide range of the parameters of the model. For
$\Omega_{m0}\!=\!0.3$, independently of the other
parameters, the model predicts that the equation of state for the dark energy today is
$w_{DE,0}\!=\!-1.4$, while $\Omega_{m0}\!=\!0.03$ leads to
$w_{DE,0}\!=\!-1.03$. In addition, during its evolution, $w_{DE}$
crosses the $w_{DE}\!=\!-1$ line to smaller values.

\end{abstract}

\maketitle

\section{Introduction}

In cosmologies where the present universe is realized as a finite
point during the cosmic evolution, the answer to the coincidence
question ``why it is that today $\Omega_{m0}$ and $\Omega_{DE,0}$ are of the same order of magnitude'',
relies on appropriate choice of \textit{initial conditions}. By contrast, in a scenario in
which the present universe is in its asymptotic era (close to a
fixed point) the answer to the above question reduces to an appropriate
choice of the \textit{parameters} of the model. However, this latter situation is not easily realized
if today's universe is accelerating, because:
\par
If the energy density of a perfect fluid with equation of state
$w\!>\!-1/3$ of any cosmological system is conserved, all fixed
points of the system with $\Omega_{m}\!\neq\! 0$ are decelerating.
\par
Indeed, with $\rho$ the energy density of the perfect fluid with conservation equation
$\dot{\rho}+3(1+w)H\rho\!=\!0$, the Hubble equation of an arbitrary cosmology
can be written in the form \be
H^{2}=2\gamma(\rho\!+\!\rho_{DE}), \label{observ}\ee where
$\gamma\!=\!4\pi G_{N}/3$. Then, the
equation governing $\rho_{DE}$ can always be brought into the form
$\dot{\rho}_{DE}\!+\!3(1\!+\!w_{DE})H\rho_{DE}\!=\!0$, where
$w_{DE}$ is time-dependent and distiguishes one model from the other. It
can be easily seen that $d(\Omega_{m}/\Omega_{DE})/d\ln
a\!=\!3(\Omega_{m}/\Omega_{DE})(w_{DE}\!-\!w)$ and
$2q\!=\!1\!+\!3(w\Omega_{m}\!+\!w_{DE}\Omega_{DE})$, where
$\Omega_{m}\!=\!2\gamma \rho/H^{2}$, $\Omega_{DE}\!=\!2\gamma
\rho_{DE}/H^{2}$ and $q\!=\!-\ddot{a}/aH^{2}$. At the fixed point (denoted by $\ast$)
$d(\Omega_{m}/\Omega_{DE})/d\ln a\!=\!0$. For  $\Omega_{m\ast}\!\neq\! 0$ one obtains
$w_{DE\ast}\!=\!w$, and $2q_{\ast}\!=\!1\!+\!3w\!>\!0$.
\par
Thus, independently of the cosmological model, the only way our accelerating universe with
$\Omega_{m\ast}\!\neq \!0$ can be close to a late time fixed point is by violating the
standard conservation equation of matter. In 4-dimensional theories, an accelerating late time
cosmological phase characterized by a frozen ratio of dark
matter/dark energy appears in coupled dark energy scenarios
\cite{amendola} as a result of the interaction of the dark
matter with other energy-momentum components, such as scalar fields. In
higher dimensional theories, where the universe is represented as a
3-brane, this violation could be the result of energy exchange
between the brane and the bulk. In particular in five dimensions, a
universe with fixed points
characterized by $\Omega_{m\ast}\!\neq\! 0$, $q_{\ast}\!<\!0$ was realized in \cite{kkttz} in the context of the
Randall-Sundrum braneworld scenario with energy influx from the bulk. However, these fixed points cannot
represent the present universe, since they have $\Omega_{m\ast}\!>\!2$. In
this paper we present a brane-bulk energy exchange model
with induced gravity whose global attractor can
represent today's universe.
\par
Let us consider an arbitrary cosmology in the form (\ref{observ}).
Instances of such cosmologies arise in braneworld models or
in theories with modified 4-dimensional actions leading to $H^{2}\!=\!f(\rho)$, or
in cosmologies where $\rho_{DE}$ is due to additional fields.
Assuming that as a result of some interaction $\rho$ is not
conserved, it will satisfy an equation of the form \be
\dot{\rho}+3(1\!+\!w)H\rho=-T. \label{oval}\ee Then, the equation
governing $\rho_{DE}$ can always be brought into the form \be
\dot{\rho}_{DE}+3(1\!+\!w_{DE})H\rho_{DE}=T,\label{obelix}\ee where
$w_{DE}$ is time and model dependent.
Whenever a fixed point of the system satisfies \be
H_{\ast}T_{\ast}\!\neq\!
0\,\,\,\,\,,\,\,\,\,\,\dot{\rho}=\dot{\rho}_{DE}=0,\label{fro}\ee
one obtains \be
w_{DE\ast}=-1-\frac{1+w}{\Omega_{m\ast}^{-1}-1}. \label{trikala}\ee
Equation (\ref{trikala}) is model-independent, in the sense that it does not
depend on the form of $T$ or the function $w_{DE}(t)$. For
$\Omega_{m\ast}\!<\!1$ equation (\ref{trikala}) gives
$w_{DE\ast}\!<\!-1$. Specifically, for $w\!=\!0$ and
$\Omega_{m\ast}\!=\!\Omega_{CDM}\!=\!0.3$ one obtains $w_{DE\ast}\!=\!-1.4$, while
for $\Omega_{m\ast}\!=\!\Omega_{bar}= 0.03$, $w_{DE\ast}\!=\!-1.03$.
\par
The cosmology discussed in the present paper has a global attractor of the form
(\ref{fro}), (\ref{trikala}) \cite{foot}. Moreover, the universe during its evolution crosses
the $w_{DE}\!=\!-1$ barrier from higher values. This behavior is favored by several recent model-independent
\cite{staro} as well as model-dependent \cite{laura,alam,leandros,phantom} analyses of the astronomical data.

\section{The model}

We consider the model described by the gravitational brane-bulk
action \cite{Dvali}
\begin{equation}
S=\int \!d^{5}x\sqrt{-g}\,(M^{3}R-\Lambda)\,+\int
\!d^{4}x\sqrt{-h}\,(m^{2}\hat{R}-V),
\end{equation}
where $R,\hat{R}$ are the Ricci scalars of the bulk metric $g_{AB}$
and the induced metric $h_{AB}\!=\!g_{AB}\!-\!n_{A}n_{B}$
respectively ($n^{A}$ is the unit vector normal to the brane and
$A,B\!=\!0,1,2,3,5$). The bulk cosmological constant is
$\Lambda/2M^{3}\!<\!0$, the brane tension is $V$, and the
induced-gravity crossover scale is $r_{c}\!=\!m^{2}/M^{3}$.
\par
We assume the cosmological bulk ansatz
\begin{equation}
ds^{2}=-n(t,y)^2dt^{2}+a(t,y)^{2}\gamma_{ij}dx^{i}dx^{j}+b(t,y)^{2}dy^{2},
\end{equation}
where $\gamma_{ij}$ is a maximally symmetric 3-dimensional metric,
parametrized by the spatial curvature $k\!=\!-1,0,1$. The non-zero
components of the five-dimensional Einstein tensor are
\begin{eqnarray}
&&\!\!\!\!\!\!G_{00}=3\Big\{\frac{\dot{a}}{a}\Big(\frac{\dot{a}}{a}\!+\!\frac{\dot{b}}{b}\Big)-
\frac{n^{2}}{b^{2}}\Big[\frac{a''}{a}\!+\!\frac{a'}{a}\Big(\frac{a'}{a}\!-\!\frac{b'}{b}\Big)\Big]+
\frac{kn^{2}}{a^{2}}\!\Big\} \!\label{eq:11}\\
&&\!\!\!\!\!\!G_{ij}\!=\!\frac{a^{2}}{b^{2}}\gamma_{ij}\Big\{\frac{a'}{a}\Big(\!\frac{a'}{a}\!
+\!\frac{2n'}{n}\!\Big)-\frac{b'}{b}\Big(\!\frac{n'}{n}\!+\!\frac{2a'}{a}\!\Big)+\frac{2a''}{a}\!+\!\frac{n''}{n}\Big\} \nn \\
&&+\frac{a^{2}}{n^{2}}\gamma_{ij}\Big\{\frac{\dot{a}}{a}\Big(\frac{2\dot{n}}{n}\!-\!
\frac{\dot{a}}{a}\Big)\!-\!\frac{2\ddot{a}}{a}\!+\!\frac{\dot{b}}{b}\Big(\frac{\dot{n}}{n}\!-\!
\frac{2\dot{a}}{a}\Big)\!-\!\frac{\ddot{b}}{b}\Big\}\!-\!k\gamma_{ij} \label{eq:12} \\
&&\!\!\!\!\!G_{05}\!=\!3\Big(\frac{n'}{n}\frac{\dot{a}}{a}+\frac{a'}{a}\frac{\dot{b}}{b}-\frac{\dot{a}'}{a}\Big)
\label{eq:13}\\
&&\!\!\!\!\!G_{55}\!=\!3\Big\{\frac{a'}{a}\Big(\!\frac{a'}{a}\!+\!\frac{n'}{n}\!\Big)-\frac{b^{2}}{n^{2}}
\Big[\frac{\ddot{a}}{a}+\frac{\dot{a}}{a}\Big(\!\frac{\dot{a}}{a}\!-\!\frac{\dot{n}}{n}\!\Big)\Big]-
\frac{kb^{2}}{a^{2}}\Big\},\! \label{eq:14}
\end{eqnarray}
where primes indicate derivatives with respect to $y$, while dots
derivatives with respect to $t$. The five-dimensional Einstein
equations take the usual form
\begin{equation}
G_{AC}=\frac{1}{2M^{3}}T_{AC}|_{tot},\label{Einstein}
\end{equation}
where \begin{eqnarray} T_{C}^{A}|_{tot}\!=
T_{C}^{A}\!|_{v,B}+T_{C}^{A}\!|_{m,B}+T_{C}^{A}\!|_{v,b}\!+T_{C}^{A}\!|_{m,b}\!+T_{C}^{A}\!|_{ind}
\end{eqnarray}
is the total energy-momentum tensor,
\begin{eqnarray}
&&\!\!\!\!\!\!\!\!\!\!\!\!T_{C}^{A}|_{v,B}=\textrm{diag}(-\Lambda,-\Lambda,-\Lambda,-\Lambda,-\Lambda)\\
&&\!\!\!\!\!\!\!\!\!\!\!\!T_{C}^{A}|_{v,b}=\textrm{diag}(-V,-V,-V,-V,0)\frac{\delta(y)}{b}\\
&&\!\!\!\!\!\!\!\!\!\!\!\!T_{C}^{A}|_{m,b}=\textrm{diag}(-\rho,p,p,p,0)\frac{\delta(y)}{b}.
\end{eqnarray}
$T_{C}^{A}|_{m,B}$ is any possible additional energy-momentum in the
bulk, the brane matter content $T_{C}^{A}|_{m,b}$ consists of a
perfect fluid with energy density $\rho$ and pressure $p$, while the
contributions arising from the scalar curvature of the brane are
given by
\begin{eqnarray}
&&\!\!\!\!\!\!\!\!T_{0}^{0}|_{ind}=\frac{6m^{2}}{n^{2}}\Big(\frac{\dot{a}^{2}}{a^{2}}+\frac{kn^{2}}{a^{2}}\Big)
\frac{\delta(y)}{b}\\
&&\!\!\!\!\!\!\!\!T_{j}^{i}|_{ind}=\frac{2m^{2}}{n^{2}}\Big(\frac{\dot{a}^{2}}{a^{2}}-\frac{2\dot{a}\dot{n}}{an}+
\frac{2\ddot{a}}{a}+\frac{kn^{2}}{a^{2}}\Big)\delta_{j}^{i}\frac{\delta(y)}{b}.
\end{eqnarray}
\par
Assuming a $\mathbb{Z}_{2}$ symmetry around the brane, the singular
part of equations (\ref{Einstein}) gives the matching conditions
\begin{equation}
\!\!\!\!\!\!\!\!\!\!\!\!\!\!\!\!\!\!\frac{a_{o^{+}}'}{a_{o}b_{o}}=-\frac{\rho\!+\!V}{12M^{3}}+\frac{r_{c}}{2n_{o}^{2}}
\Big(\frac{{\dot{a}_{o}}^{2}}{a_{o}^{2}}\!+\!\frac{kn_{o}^{2}}{a_{o}^{2}}\Big)\label{eq:15}
\end{equation}
\begin{equation}
\frac{n_{o^{+}}'}{n_{o}b_{o}}\!=\!\frac{2\rho\!+\!3p\!-\!V}{12M^{3}}
+\frac{r_{c}}
{2n_{o}^{2}}\Big(\frac{2{\ddot{a}_{o}}}{a_{o}}-\frac{{\dot{a}_{o}}^{2}}{a_{o}^{2}}-
\frac{2{\dot{a}_{o}}{\dot{n}_{o}}}
{a_{o}n_{o}}-\frac{kn_{o}^{2}}{a_{o}^{2}}\Big)\label{eq:16}
\end{equation}
(the subscript o denotes the value on the brane), while from the 05,
55 components of equations (\ref{Einstein}) we obtain
\begin{equation} \label{eq:17}
\frac{n'_{o}}{n_{o}}\frac{\dot{a}_{o}}{a_{o}}+\frac{a'_{o}}{a_{o}}\frac{\dot{b}_{o}}{b_{o}}-
\frac{\dot{a}'_{o}}{a_{o}}=\frac{T_{05}}{6M^{3}}
\end{equation}
\begin{equation} \label{eq:18}
\frac{a'_{o}}{a_{o}}\Big(\!\frac{a'_{o}}{a_{o}}+\frac{n'_{o}}{n_{o}}\!\Big)-\frac{b_{o}^{2}}{n_{o}^{2}}
\Big[\frac{\ddot{a}_{o}}{a_{o}}
+\frac{\dot{a}_{o}}{a_{o}}\Big(\!\frac{\dot{a}_{o}}{a_{o}}\!-\!\frac{\dot{n}_{o}}{n_{o}}\!\Big)
\Big]-\frac{kb_{o}^{2}}{a_{o}^{2}}\!=\!\frac{T_{55}\!-\!\Lambda
b_{o}^{2}}{6M^{3}}\!,
\end{equation}
where $T_{05},T_{55}$ are the $05$ and $55$ components of
$T_{AC}|_{m,B}$ evaluated on the brane. Substituting the expressions
(\ref{eq:15}), (\ref{eq:16}) in equations (\ref{eq:17}),
(\ref{eq:18}), we obtain the semi-conservation law and the
Raychaudhuri equation
\begin{equation}
\dot{\rho}+3\frac{\dot{a}_{o}}{a_{o}}(\rho+p)=-\frac{2n_{o}^{2}}{b_{o}}T_{5}^{0}
\label{eq:22}
\end{equation}
\begin{eqnarray}
&&\!\!\!\!\!\!\!\!\!\Big(\!H_{o}^{2}\!+\!\frac{k}{a_{o}^{2}}\!\Big)
\Big[1\!-\!\frac{r_{c}^{2}(\rho\!+\!3p\!-\!2V)}{24m^{2}}\Big]
\!+\!\frac{r_{c}^{2}(\rho\!+\!3p\!-\!2V)(\rho\!+\!V)}{144m^{4}}\nn\\
&&\!\!\!\!\!\!\!\!\!+\Big(\!\frac{\dot{H}_{o}}{n_{o}}\!+\!H_{o}^{2}\!\Big)\Big[1\!-\!\frac{r_{c}^{2}}{2}
\Big(\!H_{o}^{2}\!+\!\frac{k}{a_{o}^{2}}\!\Big)\!+\!\frac{r_{c}^{2}(\rho\!+\!V)}{12m^{2}}\Big]
\!=\!\frac{\Lambda\!-\!T_{5}^{5}}{6M^{3}}, \label{eq:19}
\end{eqnarray}
where $H_{o}\!=\!\dot{a}_{o}/a_{o}n_{o}$ is the Hubble parameter of
the brane. One can easily check that in the limit $m\! \rightarrow\!
0$, equation (\ref{eq:19}) reduces to the corresponding second order
equation of the model without $\hat{R}$ \cite{kkttz}. Energy exchange between the brane and the bulk has also
been investigated in \cite{hall, hebecker, tetra}.
\par
Since only the 55 component of $T_{AC}|_{m,B}$ enters equation
(\ref{eq:19}), one can derive a cosmological system that is largely
independent of the bulk dynamics, if at the position of the brane
the contribution of this component relative to the bulk vacuum
energy is much less important than the brane matter relative to the
brane vacuum energy, or schematically
\begin{equation}
\Big|\frac{T^{5}_{5}}{\Lambda}\Big| \ll \Big|\frac{\rho}{V}\Big|.
\label{relation}\end{equation} Then, for $|\Lambda|$ not much larger
than the Randall-Sundrum value $V^{2}/12M^{3}$, the term $T^{5}_{5}$
in equation (\ref{eq:19}) can be ignored. Alternatively, the term
$T^{5}_{5}$ can be ignored in equation (\ref{eq:19}) if simply
\begin{equation}
\Big|\frac{T^{5}_{5}}{\Lambda}\Big| \ll 1.
\label{relation1}\end{equation} Note that relations (\ref{relation})
and (\ref{relation1}) are only boundary conditions for $T_{5}^{5}$,
which in a realistic description in terms of bulk fields will be
translated into boundary conditions on these fields. In the special
case where (\ref{relation}), (\ref{relation1}) are valid throughout
the bulk, the latter remains unperturbed by the exchange of energy
with the brane.
\par
One can now check that a first integral of equation (\ref{eq:19}) is
\begin{eqnarray}
&&\!\!\!\!\!\!\!\!\!\!\!\!\!\!\!
H_{o}^{4}-\frac{2H_{o}^{2}}{3}\Big(\frac{\rho\!+\!V}{2m^{2}}\!+\!\frac{6}{r_{c}^{2}}\!-\!\frac{3k}{a_{o}^{2}}\Big)
+\Big(\!\frac{\rho\!+\!V}{6m^{2}}\!-\!\frac{k}{a_{o}^{2}}\!\Big)^{2}\!+ \nn \\
&&\,\,\,\,\,\,\,\,\,\,\,\,\,\,\,\,\,\,\,\,\,\,\,\,\,\,\,\,\,\,\,\,\,\,\,\,
+\frac{4}{r_{c}^{2}}\Big(\!\frac{\Lambda}{12M^{3}}\!-\!\frac{k}{a_{o}^{2}}\!\Big)-\frac{\chi}{3r_{c}^{2}}=0,
\label{eq:20}
\end{eqnarray}
with $\chi$ satisfying
\begin{equation}
\dot{\chi}+4n_{o}H_{o}\chi=\frac{r_{c}^{2}n_{o}^{2}\,T}{m^{2}b_{o}}\Big(\!H_{o}^{2}
\!-\!\frac{\rho\!+\!V}{6m^{2}}\!+\!\frac{k}{a_{o}^{2}}\!\Big),
\label{eq:21}\end{equation} and $T\!=\!2T_{5}^{0}$ is the
discontinuity across the brane of the 05 component of the bulk
energy-momentum tensor. The solution of (\ref{eq:20}) for $H_{o}$ is
\begin{equation} \label{eq:23}
H_{o}^{2}=\frac{\rho\!+\!V}{6m^{2}}\!+\!\frac{2}{r_{c}^{2}}\!-\!\frac{k}{a_{o}^{2}}\pm
\frac{1}{\sqrt{3}r_{c}}
\Big[\frac{2(\rho\!+\!V)}{m^{2}}\!+\!\frac{12}{r_{c}^{2}}\!-\!\frac{\Lambda}{M^{3}}\!+\!\chi\Big]^{\!\frac{1}{2}}\!,
\end{equation}
and equation (\ref{eq:21}) becomes
\begin{equation} \label{eq:24}
\dot{\chi}+4n_{o}H_{o}\chi\!=\!\frac{2n_{o}^{2}\,T}{m^{2}b_{o}}\!\Big\{\!1\pm
\frac{r_{c}}{2\sqrt{3}}
\Big[\frac{2(\rho\!+\!V)}{m^{2}}\!+\!\frac{12}{r_{c}^{2}}\!-\!\frac{\Lambda}{M^{3}}\!+\!\chi\Big]^{\!\frac{1}{2}}
\!\Big\}\!.
\end{equation}
\par
At this point we find it convenient to employ a coordinate frame in
which $b_{o}\!=\!n_{o}\!=\!1$ in the above equations. This can be
achieved by using Gauss normal coordinates with $b(t,z)\!=\!1$, and
by going to the temporal gauge on the brane with $n_{o}\!=\!1$. It
is also convenient to define the parameters
\begin{eqnarray}
\lambda & = & \frac{2V}{m^{2}}+\frac{12}{r_{c}^{2}}-\frac{\Lambda}{M^{3}} \\
\mu & = & \frac{V}{6m^{2}}+\frac{2}{r_{c}^{2}} \\
\gamma & = & \frac{1}{12m^{2}} \\
\beta & = & \frac{1}{\sqrt{3}r_{c}}.
\end{eqnarray}
For a perfect fluid on the brane with equation of state $p=w\rho$
our system is described by equations (\ref{eq:22}), (\ref{eq:23}),
(\ref{eq:24}), which simplify to (we omit the subscript o in the
following)
\begin{eqnarray}
&&\,\,\,\,\,\,\,\,\,\,\,\,\dot{\rho}+3(1+w)H\rho=-T \label{eq:27}\\
&&\!\!\!\!\!\!\!\!H^{2}=\mu+2\gamma \rho \pm
\beta\sqrt{\lambda\!+\!24\gamma
\rho\!+\!\chi}-\frac{k}{a^{2}}\label{eq:25}\\
&&\!\!\!\!\!\!\!\!\!\dot{\chi}+4H\chi=24\gamma T\Big(1\pm
\frac{1}{6\beta}\sqrt{\lambda\!+\!24\gamma
\rho\!+\!\chi}\Big)\label{eq:28},
\end{eqnarray}
while the second order equation (\ref{eq:19}) for the scale factor becomes
\begin{equation}
\frac{\ddot{a}}{a}=\mu-(1\!+\!3w)\gamma \rho \pm
\beta\frac{\lambda+6(1\!-\!3w)\gamma \rho}{\sqrt{\lambda+24\gamma
\rho+\chi}}\label{eq:26}.
\end{equation}
Finally, setting $\psi \equiv \sqrt{\lambda+24\gamma \rho+\chi}$,
equations (\ref{eq:25}), (\ref{eq:28}), (\ref{eq:26}) take the form
\begin{eqnarray}
&&\,\,\,\,\,\,\,\,\,\,\,H^{2}=\mu+2\gamma \rho \pm \beta
\psi-\frac{k}{a^{2}}
\label{eq:31}\\
&&\!\!\!\!\!\!
\dot{\psi}+2H\Big(\!\psi-\frac{\lambda+6(1\!-\!3w)\gamma
\rho}{\psi}\!\Big)=\pm \frac{2\gamma T}{\beta}
\label{eq:35}\\
&&\!\!\!\!\!\!\!\frac{\ddot{a}}{a}=\mu-(1\!+\!3w)\gamma \rho \pm
\beta \frac{\lambda+6(1\!-\!3w)\gamma \rho}{\psi}. \label{eq:32}
\end{eqnarray}
Throughout, we will assume $T(\rho)\!=\!A\rho^{\nu}$, with
$\nu>0,\,A$ constant parameters \cite{kkttz, kiritsis}. Notice that
the system of equations (\ref{eq:27})-(\ref{eq:28}) has the
influx-outflow symmetry $T\rightarrow -T$, $H\rightarrow -H$,
$t\rightarrow -t$. For $T=0$ the system reduces to the cosmology studied in \cite{Deffayet}.
\par
We will be referring to the upper (lower) $\pm$ solution as Branch A
(Branch B). We shall be interested in a model that reduces to the
Randall-Sundrum vacuum in the absence of matter, i.e. it has
vanishing effective cosmological constant. This is achieved for
$\mu\!=\!\mp\beta\sqrt{\lambda}$, which, given that
$m^{2}V\!+\!12M^{6}$ is negative (positive) for branches A (B), is
equivalent to the fine-tuning $\Lambda\!=\!-V^{2}/12M^{3}$. Notice
that for Branch A, $V$ is necessarily negative. Cosmologies with negative brane tension in the induced
gravity scenario have also been discussed in \cite{yuri}.
\par
Consider the case $k=0$. The system possesses the obvious fixed
point ($\rho_{*}, H_{*}, \psi_{*})=(0, 0, \sqrt{\lambda})$. However,
for $sgn(H) T<0$ there are non-trivial
fixed points, which are found by setting $\dot{\rho}=\dot{\psi}=0$ in
equations (\ref{eq:27}), (\ref{eq:35}). For $w\leq 1/3$ these are:
\begin{eqnarray}
&&\!\!\!\!\!\!\!\!\!\!\!\!\!\frac{2T(\rho_{*})^{2}}{9(1\!+\!w)^{2}\rho_{*}^{2}}=2\mu+(1\!-\!3w)\gamma\rho_{*}\nn\\
&&\,\,\,\,\,\pm
\sqrt{9(1\!+\!w)^{2}\gamma^{2}\rho_{*}^{2}+4\beta^{2}[\lambda+6(1\!-\!3w)\gamma\rho_{*}]}
\label{fp1}\\
&&\,\,\,\,\,\,\,\,\,\,\,\,\,\,\,\,\,\,\,\,\,\,\,\,H_{*}=-\frac{T(\rho_{*})}{3(1\!+\!w)\rho_{*}}\label{fp2}\\
&&\!\!\!\!\!\!\!\!\!\psi_{*}^{2}\pm
\frac{3(1\!+\!w)}{\beta}\gamma\rho_{*}\psi_{*}-[\lambda+6(1\!-\!3w)\gamma\rho_{*}]=0\label{fp3}.
\end{eqnarray}
Equation (\ref{eq:32}) gives
\begin{eqnarray}
&&\!\!\!\!\!\!\!\!\!\!\!\!\!\!\!\!\!\!\Big(\frac{\ddot{a}}{a}\Big)_{*}\!\!=
\frac{T(\rho_{\ast})^{2}}{9(1\!+\!w)^{2}\rho_{\ast}^{2}}\,,
\label{accel}
\end{eqnarray}
which is positive, and also, it has the same form (as a function of
$\rho_{\ast}$) as in the absence of $\hat{R}$. The deceleration
parameter is found to have the value \bea q_{\ast}=-1,
\label{decel}\eea which means $\dot{H}_{\ast}\!=\!0$. Furthermore,
at this fixed point we find
\begin{eqnarray}
\Omega_{m*}\equiv\frac{2\gamma\rho_{\ast}}{H_{\ast}^{2}}=\frac{18(1\!+\!w)^{2}}{A^{2}}\gamma\rho_{\ast}^{3-2\nu}.
\label{flat1}
\end{eqnarray}
Equation (\ref{fp1}), when expressed in terms of $\Omega_{m\ast}$,
has only one root for each branch \bea
\rho_{\ast}=\frac{\beta}{2\gamma}\frac{6(1\!-\!3w)\beta\pm\sqrt{\lambda}(1\!-\!3w\!-\!4\Omega_{m\ast}^{-1})}
{(2\Omega_{m\ast}^{-1}\!+\!1\!+\!3w)(\Omega_{m\ast}^{-1}\!-\!1)}.
\label{star} \eea However, it can be seen from (\ref{star}) that for
$-1\leq w \leq 1/3$ and $\Omega_{m\ast}< 1$ the Branch B is
inconsistent with equation (\ref{fp1}). On the contrary, Branch A
with $-1\leq w \leq 1/3$ and $\Omega_{m\ast}< 1$ is consistent for
$0<6(1\!-\!3w\!)\beta\!+\!\sqrt{\lambda}(1\!-\!3w\!-\!4\Omega_{m\ast}^{-1})
<3\sqrt{4(1\!-\!3w\!)^{2}\beta^{2}\!-\!(1\!+\!w\!)^{2}\lambda}$.
Thus, since we are interested in realizing the present universe as a
fixed point, Branch B should be rejected, and from now on we will
only consider Branch A. So, we have seen until now that {\textit{for
negative brane tension, we can have a fixed point of our model with
acceleration and $0<\Omega_{m\ast}<1$}}. This behavior is
qualitatively different from the one obtained in the context of the
model presented in \cite{kkttz} (for $-1/3 \! \leq\!w\!\leq\!1/3$),
where for positive brane tension we have $\Omega_{m\ast}>2$, while
for negative brane tension the universe necessarily exhibited
deceleration; therefore, in that model the idea that the present
universe is close to a fixed point could not be realized.
\par
Concerning the negative brane tension the following remarks are in
order: (a) In the conventional, non-supersymmetric setting,
it is well known that a negative tension brane with or without
induced gravity is accompanied by tachyonic bulk gravitational modes
\cite{ratazi}; however, including the
Gauss-Bonnet corrections relevant at high-energies, the tachyonic
modes can be completely removed for a suitable range of the
parameters \cite{charmousis}. (b) As shown in \cite{stelle}, in
supersymmetric theories, spacetimes with two branes of opposite
tension are stable; in particular, there is no instability due to
expanding ``balooning'' modes on the negative brane. It is, however,
unclear what happens in models with supersymmetry unbroken in the
bulk but softly broken on the brane. (c) Finally, it has been shown \cite{smolyakov} that with appropriate
choice of boundary conditions, both at the linearized level as well as in the full theory,
the gravitational potential of a mass on a negative tension brane has the correct $1/r$ attractive behaviour.

\section{Critical point analysis}

We shall restrict ourselves to the flat case $k\!=\!0$. In order to study the dynamics of the system,
it is convenient to use (dimensionless) flatness parameters such that the state space is
compact \cite{goheer}. Defining
\be \omega_{m}\!=\!\frac{2\gamma\rho}{D^{2}}
\,\,\,\,\,,\,\,\,\,\,\omega_{\psi}=\frac{\beta\psi}{D^{2}}\,\,\,\,\,,\,\,\,\,\,
Z=\frac{H}{D}\,, \label{flat} \ee where
$D\!=\!\sqrt{H^{2}\!-\!\mu}$, we obtain the equations
\begin{eqnarray}
&&\,\,\,\,\,\,\,\,\,\,\,\,\,\,\,\,\,\,\,\,\,\,\,\,\,\,\,\,\,\,\,\,\,\,\,\,\,\,\,\,\,\,\,
\omega_{m}+\omega_{\psi}=1
\label{friflat}\\
&&\!\!\!\!\!\!\!\omega_{m}'\!=\!\omega_{\!m}\!\Big[\!(1\!+\!3w)(\omega_{\!m}\!\!-\!1\!)Z\!-\!\frac{
A}{\sqrt{|\mu|}}
\Big(\!\frac{|\mu|\omega_{\!m}}{2\gamma}\!\Big)^{\!\!\nu\!-\!1}\!(1\!-\!Z^{2})^{\frac{3}{2}-\nu}\nn\\
&&\,\,\,\,\,\,\,\,\,\,\,\,\,\,\,\,\,\,\,\,\,\,\,
-2Z(1\!-\!Z^{2})\frac{1\!-\!Z^{2}\!-\!3(1\!-\!3w)\beta^{2}\mu^{-1}\omega_{m}}
{1\!-\!\omega_{m}}\!\Big] \label{gerold}\\
&&\!\!\!\!\!\!\!Z'\!=\!(1\!-\!Z^{2})\Big[(1\!-\!Z^{2})
\frac{1\!-\!Z^{2}\!-\!3(1\!-\!3w)\beta^{2}\mu^{-1}\omega_{m}}{1\!-\!\omega_{m}}-1\nn\\
&& \,\,\,\,\,\,\,\,\,\,\,\,\,\,\,\,\,\,\,\,\,\,\,\,
\,\,\,\,\,\,\,\,\,\,\,\,\,\,\,\,\,\,\,\,\,\,\,\,\,\,\,\,\,\,\,\,\,\,\,\,\,\,\,\,\,\,\,\,\,\,\,\,
\,\,\,\,\,\,\,\,\,\,\,\,\,\,\,\,\,\,\,\,-\frac{1\!+\!3w}{2}\omega_{m}\Big],
\label{italy}
\end{eqnarray}
with $'\!=\!d/d\tau\!=\!D^{-1}d/dt$. Note that $-1\leq Z\leq 1$,
while both $\omega$'s satisfy $0\leq \omega\leq 1$. The
deceleration parameter is given by \be
q\!=\!\frac{1}{Z^{2}}\!\Big[\!\frac{1\!+\!3w}{2}\omega_{m}\!-\!(1\!-\!Z^{2})
\frac{\omega_{m}\!-\!\!Z^{2}\!-\!3(1\!-\!3w)\beta^{2}\mu^{-1}\omega_{m}}{1\!-\!
\omega_{m}}\!\Big]\label{greece} \ee and
$H'=-HZ(q+1)$. The system of equations
(\ref{gerold})-(\ref{italy}) inherits from equations
(\ref{eq:27})-(\ref{eq:28}) the symmetry $A\rightarrow -A$,
$Z\rightarrow -Z$, $\tau\rightarrow -\tau$. The system written in the new variables contains only three
parameters. However, going back to the physical quantities $H$, $\rho$ one will need specific values of
two more parameters.
\par
It is obvious that the points with $|Z|=1$ have $H=\infty$.
Therefore, from (\ref{eq:31}) it arises that the infinite
density $\rho\!=\!\infty$ big bang (big crunch)
singularity, when it appears, is represented by one of the points with
$Z\!=\!1$ ($Z\!=\!-1$). The points
with $\omega_{m}\!=\!1$, $|Z|\!\neq\! 1,0$ have $\omega_{m}'\!=\infty$,
$Z'\!=\infty$ and finite $\rho$, $H$; for
$w\!\leq \!1/3$, one has in addition $\ddot{a}/a\!=\!+\infty$, i.e.
divergent 4D curvature scalar on the brane.
\par
The system possesses, generically, the fixed point (a)
$(\omega_{m\ast},\omega_{\psi\ast},Z_{\ast})\!=\!(0,1,0)$, which
corresponds to the fixed point
$(\rho_{\ast},H_{\ast},\psi_{\ast})\!=\!(0,0,\sqrt{\lambda})$ discussed above. For $\nu\!\leq\! 3/2$
there are in addition the fixed points (b)
$(\omega_{m\ast},\omega_{\psi\ast},Z_{\ast})\!=\!(0,1,1)$ and (c)
$(\omega_{m\ast},\omega_{\psi\ast},Z_{\ast})\!=\!(0,1,-1)$. All these
critical points are either non-hyperbolic, or their characteristic
matrix is not defined at all; thus, their stability cannot be
studied by first order perturbation analysis. In cases like these, one can find non-conventional
behaviors (such as saddle-nodes and cusps \cite{perko}) of the flow-chart near the critical points.
There are two more candidate fixed points (d)
$(\omega_{m\ast},\omega_{\psi\ast},Z_{\ast})=(1,0,1)$ and (e)
$(\omega_{m\ast},\omega_{\psi\ast},Z_{\ast})=(1,0,-1)$, whose
existence cannot be confirmed directly from the dynamical system,
since they make equations (\ref{gerold}), (\ref{italy})
undetermined. Apart from the above, there are other critical points given by \bea
&&\,\,\,\,\,\,\,\,\,\,\,\,\,\,\,\,\,\,\,\,
\frac{A}{\sqrt{|\mu|}}\Big(\!\frac{|\mu|\omega_{m\ast}}{2\gamma}\!\Big)^{\!\nu-1}\!=
-\frac{3(1\!+\!w)\,Z_{\ast}}{(1\!-\!Z_{\ast}^{2}) ^{\frac{3}{2}-\nu}}\label{jack}\\
&&\!\!\!\!\!\!\!
(1\!+\!3w\!)\omega_{\!m\ast}^{2}\!\!+\!(1\!-\!3w\!)\Big[\!1\!-\!\frac{6\beta^2}{\mu}
\!(1\!-\!Z_{\ast}^{2})\!\Big]\omega_{\!m\ast}\!\!-\!2[1\!-\!(1\!-\!Z_{\ast}^{2})^{2}]\nn\\
&&\,\,\,\,\,\,\,\,\,\,\,\,\,\,\,\,\,\,\,\,\,\,\,\,\,\,\,\,\,\,\,\,\,\,\,\,\,\,\,\,\,\,\,\,\,\,\,\,\,
\,\,\,\,\,\,\,\,\,\,\,\,\,\,\,\,\,\,\,\,\,\,\,\,\,\,\,\,\,\,\,\,\,\,\,\,\,\,\,\,\,\,\,\,\,\,\,\,\,\,
\,\,\,\,\,\,\,\,\,\,\,\,\,\,\,\,\,=\!0.\label{office} \eea They exist only for $A Z_{\ast}\!<\!0$ and
correspond to the ones given by equations
(\ref{fp1})-(\ref{fp3}). For the physically interesting case $w\!=\!0$ with influx we scanned the parameter
space and were convinced that for $\nu\!\neq\!3/2$ there is always only one fixed point; for $\nu\!<\!3/2$
this is an attractor ($\textsf{A}$), while for $\nu\!>\!3/2$ this is a saddle ($\textsf{S}$).
For $w\!=\!0$, $\nu\!=\!3/2$ there is either one fixed point (attractor) or no fixed points, depending on the
parameters. For the other characteristic value $w\!=\!1/3$, we concluded that for $\nu\!<\!3/2$ there is only one
fixed point (attractor), for $\nu\!>\!2$ there is only one fixed point (saddle), while for $3/2\!<\!\nu\!<\!2$
there are either two fixed points (one attractor and one saddle) or no fixed points at all, depending on the
parameters. For $w\!=\!1/3$, $\nu\!=\!3/2$ there is either one fixed point (attractor) or no fixed points.
Finally, for $w\!=\!1/3$, $\nu\!=\!2$ there is either one fixed point (saddle) or no fixed points. These results
were obtained numerically for a wide range of parameters and are summarized in Tables 1 and 2.
\vspace{0.3cm}
\begin{center}
\begin{tabular}{|c|c|c|c|}
\hline
 & $\nu<3/2$ & $\nu=3/2$ & $\nu>3/2$  \\ \hline
 No. of F.P. & 1 & 0 or 1 & 1 \\ \hline
 Nature & \textsf{A} & \,\,\,\,\,\,\,\,\,\,\,\,\textsf{A} & \textsf{S} \\
\hline
\multicolumn{4}{l}{Table 1: The fixed points for w=0, influx}
\end{tabular}
\end{center}
\vspace{0.05cm}
\begin{center}
\begin{tabular}{|c|c|c|c|c|c|}
\hline
 & $\nu\!<\!3/2$ & $\nu\!=\!3/2$ & $3/2\!<\!\nu\!<\!2$ & $\nu\!=\!2$ & $\nu\!>\!2$  \\ \hline
 No. of F.P. & 1 & 0 or 1 & \!\!0 or 2 & 0 or 1 & 1\\ \hline
 Nature & \textsf{A} & \,\,\,\,\,\,\,\,\,\,\,\,\textsf{A} & \,\,\,\,\,\,\,\,\,\,
 \textsf{A},\textsf{S} & \,\,\,\,\,\,\,\,\,\,\,\textsf{S} & \textsf{S} \\
\hline
\multicolumn{6}{l}{\,\,\,\,\,\,\,\, Table 2: The fixed points for w=1/3, influx}
\end{tabular}
\end{center}
\vspace{0.15cm}
The approach to an attractor described by the linear approximation of (\ref{gerold})-(\ref{italy})
is exponential in $\tau$ and takes infinite time $\tau$ for the universe to reach it. Given that near
this fixed point the relation between the cosmic time $t$ and the
time $\tau$ is linear, we conclude that it also takes infinite
cosmic time to reach the attractor.
\par
Defining $\epsilon\!=\!sgn(H)$, we see from (\ref{gerold})-(\ref{italy}) that the lines $Z \!=\! \epsilon$ ($\nu\!\leq\! 3/2$),
$\omega_{m}\!=\! 0$ are orbits of the system. Furthermore, the
family of solutions with $Z\!\approx \!\epsilon$ and
$dZ/d\omega_{m}\!=\!Z'/\omega_{m}'\!\approx\! 0$ is approximately
described for $\nu\!<\!3/2$ by
$\omega_{m}'\!=\!\epsilon(1\!+\!3w)\omega_{m}(\omega_{m}\!-\!1)$, and thus, they move away from the point
$(\omega_{m\ast},Z_{\ast})\!=\!(1,1)$, while they approach the point
$(\omega_{m\ast},Z_{\ast})\!=\!(1,-1)$. In addition, the solution of this
equation is
$\omega_{m}\!=\![1\!+\!ce^{\epsilon(1\!+\!3w)\tau}]^{-1}$, with
$c\!>\!0$ an integration constant. Using this solution in
equation $H'/H\!=\!-Z (q\!+\!1)$ we find that for $w\!=\!1/3$,
$H/H_{o}\!=\!\sqrt{\omega_{m}}/(1\!-\!\omega_{m})$, where $H_{o}$
is another integration constant. Then, the equation for $\omega_{m}(t)$ becomes
$d\omega_{m}/dt\!=\!-2\epsilon\omega_{m}\sqrt{H_{o}^{2}\omega_{m}\!-\!\mu(1\!-\!\omega_{m})^{2}}$,
and can be integrated giving $t$ as a function of $\omega_{m}$ or
$H$. Therefore, in the region of the big bang/big crunch singularity
one obtains $a(t)\!\sim \!\sqrt{\epsilon t}$, $\rho(t)\!\sim\!t^{-2}$, as in the standard radiation
dominated big-bang scenario. This means that for $\nu\!<\!3/2$ the energy exchange has no observable effects
close to the big bang/big crunch singularity.
\begin{figure}[h!]
\centering
\begin{tabular}{cc}
\includegraphics*[width=240pt, height=180pt]{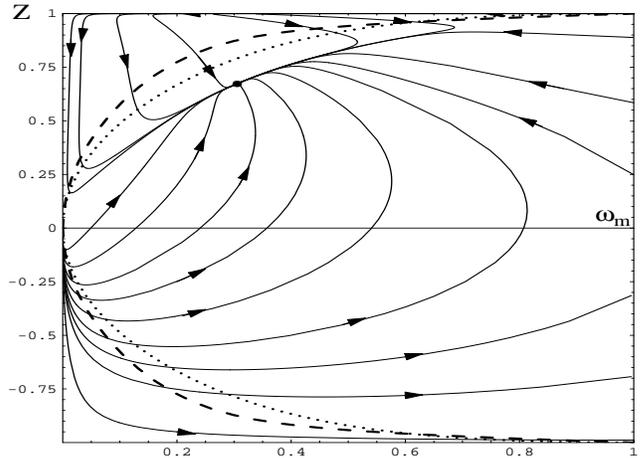}&%
\end{tabular}
 \caption{
Influx, $w\!=\!0$, $\nu\!<\!3/2$. The arrows show the
direction of increasing cosmic time. The dotted line
corresponds to $w_{DE}\!=\!-1$. The region inside (outside) the dashed line corresponds to acceleration
(deceleration). The region with $Z\!>\!0$ represents expansion, while
$Z\!<\!0$ represents collapse. The present universe is supposed to be close to the global attractor.}
\end{figure}
\begin{figure}[h!]
\centering
\begin{tabular}{cc}
\includegraphics*[width=240pt, height=180pt]{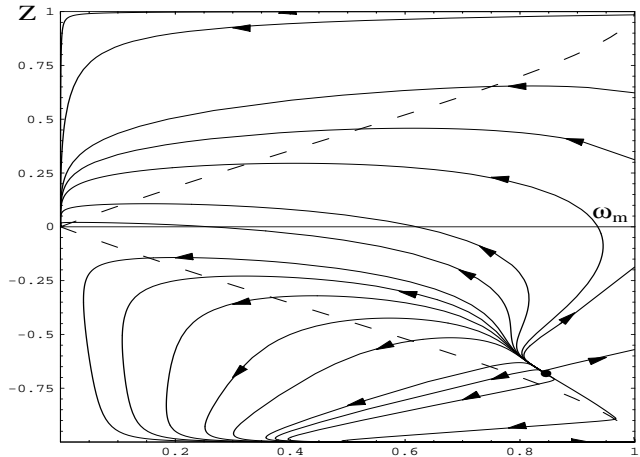}&%
\end{tabular}
 \caption{
Outflow, $w\!=\!1/3$, $\nu\!<\!3/2$. The arrows show the
direction of increasing cosmic time. The region inside (outside) the dashed line corresponds to acceleration
(deceleration). The region with $Z\!>\!0$ represents expansion, while
$Z\!<\!0$ represents collapse.}
\end{figure}
\par
Since our proposal relies on the existence of an attractor, we shall restrict ourselves to the case
$\nu\!<\!3/2$. It is convenient to discuss the four possible cases separately:
\newline
(i) $w\!=\!0$ with influx. The generic behavior of the solutions of equations
(\ref{gerold})-(\ref{italy}) is shown in Figure 1. We see that all the expanding solutions approach the global attractor.
Furthermore, there is a class of collapsing solutions which bounce to expanding ones.
Finally, there are solutions which collapse all during their lifetime to a state with finite $\rho$ and $H$.
The physically interesting
solutions are those in the upper part of the diagram emanating from the big bang $(\omega,Z)\!\approx\!(1,1)$.
These solutions start with a period of deceleration. The subsequent evolution depends on the value of
$3\beta^{2}\!/|\mu|$, which determines the relative position of the dashed and dotted lines.
Specifically, for $3\beta^{2}\!/|\mu|\!>\!1$ (the case of Figure 1) one distinguishes two possible
classes of universe evolution. In the first, the universe crosses the dashed line entering the
acceleration era still with $w_{DE}\!>\!-1$, and finally it crosses the dotted line to $w_{DE}\!<\!-1$ approaching
the attractor. In the second, while in the deceleration era,
it first crosses the dotted line to $w_{DE}\!<\!-1$, and then the dashed line entering the
eternally accelerating era. For $3\beta^{2}\!/|\mu|\!\leq\!1$, the dotted line lies above the dashed line,
and, consequently, only the second class of trajectories exists. To connect with the discussion in the introduction, notice that the
Friedmann equation (\ref{eq:31}) can be written in the form
(\ref{observ}) with dark energy
$\rho_{DE}\!=\!(\beta\psi\!+\!\mu)/2\gamma$. Using (\ref{eq:35}), the equation for $\rho_{DE}$ takes the form
(\ref{obelix}) with \be
\!w_{DE}\!=\!\frac{-1}{3(1\!-\!\omega_{m})}\Big[\!2Z^{2}\!-\!\omega_{m}\!-\!1\!-\!6(1\!-\!3w)\frac{\beta^{2}}{\mu}
\frac{\omega_{m}(1\!-\!Z^{2})}{Z^{2}\!-\!\omega_{m}}\!\Big]\!.
\label{dark}\ee  The global attractor (\ref{fp1})-(\ref{fp3})
satisfies relations (\ref{fro}) and consequently, $w_{DE}$ evolves to the value $w_{DE\ast}$ given
by (\ref{trikala}). As for the bouncing solutions, they approach the attractor
after they cross the line
$Z^{2}\!=\!\omega_{m}$, where $w_{DE}$ jumps from $+\infty$ to $-\infty$; however,
the evolution of the observable quantities is regular.
\newline
(ii) $w\!=\!0$ with outflow. The generic behavior in this case is obtained from Figure 1
by the substitution $Z\!\rightarrow\! -Z$ and $\tau\rightarrow -\tau$, which reflects the diagram
with respect to the $\omega_{m}$ axis and converts attractors to repelers.
\newline
(iii) $w\!=\!1/3$ with outflow. Figure 2 depicts the flow diagram of this case. Even though in the case of
radiation in general $w_{DE}\!>\!-1/3$ from equation (\ref{dark}), there are both acceleration and deceleration
regions. Furthermore, from equation (\ref{trikala}) it is
$\Omega_{m\ast}\!>\!1$.
\newline
(iv) $w\!=\!1/3$ with influx. This arises like in (ii) by reflection of Figure 2 and resembles Figure 1.

\section{Conclusions}

In this work, we studied the role of brane-bulk energy exchange on the cosmological evolution
of a brane with negative tension, zero effective cosmological constant, and in the presence of the
induced curvature scalar term in the action. Adopting the physically motivated $\rho^{\nu}$ power-law form
for the energy transfer and assuming a cosmological constant in the bulk, an autonomous system of equations
was isolated. In this scenario, the ``dark energy'' is a result of the geometry and the brane-bulk
energy exchange. The negative tension of the brane is necessary in order to
realize the present universe (accelerating with $0\!<\!\Omega_{m0}\!<\!1$) as being close to a future fixed point
of the evolution equations. We studied the possible cosmologies using bounded normalized variables and the
corresponding global phase portraits were obtained.
By studying the number and nature of the fixed points we demonstrated numericaly that our present universe
can be easily realized as a late-time fixed point of the evolution. This provides an alternative answer
to the coincidence problem in cosmology, which does not require specific fine-tuning of the
initial data. Furthermore, the equation of state for the dark energy at the attractor is uniquely
specified by the value $\Omega_{m0}$. Remarkably, for
$\Omega_{m0}\!=\!0.3$, one obtains $w_{DE,0}\!=\!-1.4$, independently of the other parameters, while for
the other suggestive value $\Omega_{m0}\!=\!0.03$, $w_{DE,0}\!=\!-1.03$.
In the past, the function $w_{DE}$ crosses the line $w_{DE}\!=\!-1$ to larger values.
\par
It would be interesting to investigate if the above
partial success of the present scenario persists after one tries to fit the supernova data
and the detailed CMB spectum \cite{miz}. Of course, the nature of the content of the bulk and
of the mechanism of energy exchange with the brane is another crucial open question, which we hope
to deal with in a future publication.

\[ \]
{\bf Acknowlegements.} Supported in part by the EU grant
MRTN-CT-2004-512194. The work of G.K. is also supported by the
European Commission Marie-Curie Fellowship under contract
MEIF-CT-2004-501432. G.P. is also supported by the Greek Ministry of
Education research program ``Herakleitos''.


\begin{references}
\bibitem{foot} This is in contrast to the models discused in \cite{amendola}
in which equation (\ref{fro}) is not satisfied.
\end{references}
\end{document}